\documentclass{cdclass}

\begin{document}

\title{Analysis of pedestrian stress level using GSR sensor in virtual immersive reality}
%\titlerunning{Short form of title} % if too long for running head

\author{
  Mahwish Mudassar\authorlabel{1} \and 
  Arash Kalatian\authorlabel{2} \and
  Bilal Farooq\authorlabel{3} 
}
%short form of author list (without footnote symbol) for running head; may be using "et al." 
\authorrunning{M. Mudassar \and A. Kalatian \and B. Farooq}
\institute{
  \authorlabel{1} Laboratory of Innovations in Transportation (LiTrans), Ryerson University, Toronto, Canada,
  \authoremail{1}{Mahwish.Mudassar@ryerson.ca}
  \and
  \authorlabel{2} Institute of Transport Studies, University of Leeds, Leeds, United Kingdom, 
  \authoremail{2}{A.Kalatian@leeds.ac.uk}
  \and
  \authorlabel{3} Laboratory of Innovations in Transportation (LiTrans), Ryerson University, Toronto, Canada, 
  \authoremail{3}{Bilal.Farooq@ryerson.ca}
}

% filled out by the editor: %%%%%%%%%%%%%%%%%%%%%%%%%%%%%%%%%%%%%%%
\date{2021}{August 15, 2021}{November 17, 2021}{November 22, 2021} % dates (received/revised/accepted)
\ldoi{10.17815/CD.20XX.X} % doi 
\volume{V}  % volume 
\online{AX} % url 
%%%%%%%%%%%%%%%%%%%%%%%%%%%%%%%%%%%%%%%%%%%%%%%%%%%%%%%%%%%%%%%%%%%

\maketitle

\begin{abstract}
Level of emotional arousal of one's body changes in response to external stimuli in an environment. Given the risks involved while crossing streets, particularly at unsignalized mid-block crosswalks, one can expect a change in the stress level of pedestrians. In this study, we investigate the levels and changes in pedestrian stress, under different road crossing scenarios in immersive virtual reality.
To measure stress level of pedestrians, we used Galvanic Skin Response (GSR) sensors.
To collect the required data for the model, Virtual Immersive Reality Environment (VIRE) \cite{farooq2018virtual} tool is used, which enables us to measure participant's stress levels in a controlled environment. The results suggested that the density of vehicles has a positive effect, meaning as the density of vehicles increases, so does the stress levels for pedestrians. It was noted that younger pedestrians have lower amount of stress when crossing as compared to older pedestrians which have higher amounts of stress. Geometric variables has an impact on the stress level of pedestrians. The greater the number of lanes the greater the observed stress, which is due the crossing distance increasing, while the walking speed remaining the same. 

\end{abstract}

% e.g. self-driven particles; pedestrians; vehicles; animals;
% molecular motors; agents; traffic; crowds; swarms; experiments;
% models; microscopic; macroscopic; celular automata; software; open
% source

%%%%%%%%%%%%%%%%%%%%%%%%%%%%%%%%%%%%%%%%%%%%%%%%%%%%%%%%%%%%%%%%%%%
\section{Introduction}
\label{sec:intro}

It is envisioned that within the next few decades connected autonomous vehicles (CAVs) will replace the standard car. The goal of CAVs is to reduce accidents caused by human error, decrease pollution, decrease congestion, and improve road traffic. Human error when driving causes 90\% of accidents, which will be significantly reduced when autonomous vehicle driven cars are used \cite{khayyam2020artificial}. Currently, the advanced features of cars (level 2 autonomous) are cruise control, parking assistance, emergency braking, lane keeping assistance. Fully autonomous vehicles can also decrease pollution and congestion, this is due to cars being able to control the speed and path they travel.

Before the ubiquitous use of fully autonomous vehicles on urban roads, there is a need for substantial research on their effects on pedestrian, cyclists, and human driven vehicles. An important research question is to understand how CAVs will influence the stress levels of pedestrian when crossing the urban road. Pedestrians walking in urban spaces should be evaluated to determine their level of comfort with CAVs when crossing signalized as well as unsignalized intersections. With the study of pedestrian stress, the adaptation of pedestrian-vehicle interactions can be recorded and evaluated. The study of pedestrians is extremely important as pedestrian movement and interaction with CAVs need to be evaluated, to incorporate safety and comfort features. Research on pedestrian-CAV interactions have started to become more prevalent in the last couple of years. These recent studies will be one of the main motivators for safety in urban environments with CAVs.

The contribution of this research is unique, as it evaluates pedestrian stress and the interactions of pedestrian with CAVs. We used immersive virtual reality and galvanic skin resistance (GSR) sensor to collect data on pedestrians in a controlled environment. The data are then analyzed to explore the effects of various vehicle type, flow, sociodemographic, and geometric variables on the stress levels. This research will benefit stakeholders, the government and research agencies as the interactions between human driven and automated cars will aid in developing new policies and considerations for CAVs in the market. For government agencies, new policies, procedures and requirements for the transition of human driven vehicles to autonomous vehicles are important for the health and safety of the community. The transition period for these vehicles will take years before automated vehicles will completely implemented in the market. For stakeholders and research agencies, this research will give an understanding of how stress impacts decision for crossing. A pattern can also be found when evaluating multiple participants on the reaction of stress as well as choices made, during periods of low, medium and high levels of stress. For researchers this research can be further developed to understand and implement the safety features for automated vehicles as well as understand the risks and liabilities for an automated vehicle.

The next section discusses previous research conducted on pedestrian crossing behaviour and stress levels. An overview of the experiment and the methodology as well as data processing is discussed. There is an in depth analysis of results and discussion which consists of a correlation analysis and microscopic and temporal analysis of the stress. Lastly, future work and research are discussed. 
\section{Literature review}
\label{sec:1}
When looking at the existing literature on pedestrian behaviour, it was noted that the data were collected using five primary methods: Virtual Reality/Simulation, Questionnaires or Surveys, Observational data, Galvanic Skin Response (GSR) and Electroencephalograms (EEG). Often simulations were paired with survey questionnaires, so that there was a better understanding of the participants and data. The topics evaluated during literature review were child safety, decision making, distracted pedestrians, effects of gender and age for pedestrians, pedestrian research and street sharing effects of safety on pedestrian choice. 

\subsection{Child Safety}
\label{sec:2}
Children are an uncommon demographic to study when determining pedestrian behaviour. This is due to children being one of the hardest demographics to study. This is due to the limits of experimentation with children. Children tend to treat experiments as a game rather than an experiment, thus they would perform risky behaviours they wouldn't otherwise. To resolve this issue, exposure to the simulation for a few minutes will help children take the experiment more seriously. This is because children will be able to explore and have fun in the simulation and it still will be a relatively new experience when conducting the experiment. It was shown in studies that children are less observant than their adult counterparts when crossing the road \cite{meir2015child}. When observing children it was discovered that, the more exposure for a particular experience a person goes through, the less likely they are stressed. Once adapted to a situation, to achieve the amount of stress before adaptation, more stimulus is needed \cite{quy1974comparison}. The stress levels of children are important to understand due to them being the most vulnerable groups to be injured by autonomous vehicles.

\subsection{Decision Making}
\label{sec:2}
Decision making is an important aspect which was researched in order to understand pedestrian stress. There are multiple factors that influence the decision making abilities of crossing. The norms of society and social pressure have an important role to determine the decisions a pedestrian makes when crossing. There are different levels of acceptability for pedestrian actions in different countries. In France, it is more acceptable to cross while the light is red than it is in Japan \cite{feliciani2020efficiently}. Due to this there is less stress and lower wait times for pedestrians in France compared to pedestrians in Japan.  Wait times impact the stress level and pedestrian decision making process. It was noted that if pedestrians wait longer than 40 second on a red signal, the pedestrian will cross the street despite the red signal \cite{bendak2021factors}. Weather is a significant factor when determining wait time and stress levels of a pedestrian. When the weather conditions are warmer, pedestrians have smaller wait times and greater levels of stress \cite{bendak2021factors}. In other studies it was found that when there were a greater number of lanes, pedestrians were more hesitant and less willing to cross the road on a red signal. The speed for walking increased when the number of lanes increase. This is due to the distance for crossing increasing thus, the walking speed would need to increase in order to cross the road with the minimum amount of time. With the increase in the number of lanes, pedestrians were noted to deeply consider crossing on a red light \cite{shaaban2017agent}. 
The decision making abilities, and stress levels are varied when there is a crowd or group present. When there is a group of pedestrians, it was noted that they are more likely to violate the regulations and rules of crossing \cite{brosseau2013impact}. This contributes to the herd mentality which is common in large groups and crowds. The herd mentality is when a group of people conform to the belief of the majority in a group. In terms of crossing behaviour this means that the “leader” of the group makes a decision the rest of the group will follow \cite{faria2010collective}. This mentality and group decision making is a result of social cues that society has formed. When examining the stress levels of the group “followers”, it was shown that they are more stressed when crossing the stress as they have not consciously made the choice of following, but rather the choice was made by the “leader” \cite{faria2010collective}. The stress levels of a group versus individual crossing are similar.
In an emergency situation decision making skills and stress levels for an individual were studied. In an emergency situation, decision making skills are influenced as there are many factors to consider in a shorter span of time. Pedestrians are known to take the closest exit route choice available to them as they are more comfortable, especially when the route is not blocked or crowded \cite{mavros2016geo}. Furthermore, it was shown that pedestrians are willing to take a further exit if it has potentially less wait time \cite{haghani2016human}. When exiting if the exist way is clear of obstructions then the stress level is relatively normal but with an obstruction the stress level increases \cite{li2017effect}. At a crossing, when pedestrians are walking with little to no room, the amount of stress increases when walking \cite{osaragi2004modeling}. This ultimately means that the more crowded an intersection the greater the stress levels.

\subsection{Gender and Age for Pedestrians}
\label{sec:2}
The effects of stress levels and wait times can be determined by gender and age. Males have different biological thinking than women \cite{kadali2012pedestrians}. The reaction to stress for males and females are also different and thus it affects the decision making process. It was studied for road crossing when compared to men, women are more hesitant for crossings \cite{holland2007effect}. As a result, women are more stressed than men when crossing. It was noted women would take their time and cross safely, even if it meant that they had to wait longer. This is due to the fact women perceive more risk when crossing as compared to men \cite{li2013pedestrian}. The social normality and stereotypes perceived for each sex also alter their view on crossing and safety. As compared to men, women are told to value rules and regulations from an early age. This is shown when girls' hands are held more often compared to boys when crossing the road \cite{zeedyk2003behavioural}. This underlies that girls are seen as needing protection more than boys, thus internally they would grow up feeling more cautious with the rules and regulations around them.   
Age is an influential factor when determining stress levels and adaptation. It was shown that senior pedestrians were more likely to underestimate the crossing time \cite{oxley2005crossing}. Therefore, seniors were more stressed out when crossing due to the timer ending before the senior had time to finish the cross. This is due to seniors underestimating their physical abilities and their limitations. The most common underestimation for seniors is gap time, due to seniors having their speeds slower and mentally assuming their speed is the same as before. For children the stress levels are higher, this is due to children not having exposure crossing the road \cite{meir2015child}. When experiencing new stimulus, the stress levels are increased, once there is more exposure the stress levels decrease \cite{quy1974comparison}. 

\subsection{Street Sharing and Pedestrian Choice}
\label{sec:2}
Pedestrians and vehicles commonly share the street with one another. Other factors such as vehicles on the road impact the stress of a pedestrian and their time of crossing. The effects of sharing the street has an effect on the stress levels. In one experiment it was shown that when pedestrians were initially reacting to connected automated vehicles (CAVs), it was difficult for them to communicate with the CAVs. The interaction and communication between AVs and pedestrians affected crossing decisions \cite{rad2020pedestrians}. Due to some miscommunications, initially with autonomous vehicles, it has caused accidents between the pedestrian and the vehicle. The interactions were different depending on age group. Seniors have different interactions and ways of communication compared to their adult and children counterparts \cite{cloutier2017outta}.
The prediction of pedestrian actions in a shared space is important to understand. This will aid in future safety measures that are in place of autonomous vehicles.  When pedestrians are told which signals to look for in a CAV and the instructions from a vehicle that is better understood, than vehicle status \cite{ackermann2019experimental}. When given directions by the vehicle, it reduced the amount of miscommunication and accidents. The communication between pedestrians and vehicles would also be different depending on the country and society, thus different communication styles would need to be taken into consideration when designing for CAVs. 

\subsection{Remarks}
The current literature related to interactions of pedestrians with CAV is limited. Some studies have collected the data using using a small number of participants interacting with a limited CAV e.g., slow moving autonomous shuttle. Stress levels were not considered rather a questionnaire was completed and adaptation to the CAVs was measured using the questionnaire. To the best of our knowledge, no research has monitored the stress level of pedestrians when interacting with a CAV. This research is unique as stress levels were measured using galvanic skin resistance sensor when interacting with AV in virtual reality and adaptation was considered when the same experiment and conditions were repeated. The change of stress levels were  analyzed from the two experiences. This study explores the effects of stress levels for crossing behaviour using GSR sensor and virtual reality.

\section{Methodology}
\label{sec:2}
The study focused on mid-block crossing behaviours at unsignalized intersections. The participant was asked to fill out a questionnaire to collect social demographic information including age, gender, occupation, education, driving license holder, number of cars in household, primary mode of transportation, number of times the participants walk, previous experience with virtual reality, if the participant struggles with vision, anxiety, headaches, and dizziness. If the participant revealed that they struggled with vision, anxiety, headaches and dizziness, they were not proceeded to the main experiment.  

The participant was then asked to wear a VR headset, and a GSR sensor on two fingers on one hand. The GSR skin conductance was used to measure the stress levels of a participant. The GSR sensor uses a small electrical charge to measure the amount of sweat an individual has on their finger. The greater the charge the greater the sweat. Sweat is a form of arousal which indicates a level of stress, the greater the sweat an individual has the greater the level of stress. Thus, the stress levels can be monitored for an individual over a period of time\cite{kalatian2021decoding}. 

The participant was shown a different street scenario in each experiment.  These experiments have multiple variables, which include speed limit, lane width, minimum gap time (minimum gap time between two vehicles), full braking before impact, mean arrival rate of vehicles, road type (two way or one way traffic), day or night, weather, and automation level.The different variables available for each experiment is shown in \tref{table:1}. Due to a large number of combinations possible, a D-optimal design was created and 86 unique combinations were identified for the experiments. There were a total of 180 participants that participated in the experiment over the course of a 5-month period. The participants were collected from four main locations in the Greater Toronto Area (GTA) to include a variety of ages and experiences.  Participants were asked to cross the road when they felt it was safe to do so and their stress level was recorded alongside the crossing. The participants x, y, and z coordinates were recorded using the VR system and python coding which helped determine their changing location. The participant's instantaneous speed and acceleration levels were evaluated and compared alongside their stress levels. The participants that were chosen for analysis based on a few conditions, such as there were no major health problems, the participant was able to complete the experiment successfully without any issues in reading the GSR sensor. For a detailed description and analysis of the data refer to \cite{kalatian2021context, kalatian2021decoding}.

% table example
\begin{table}
\small
  \centering
  \begin{tabular}{llll}
    \toprule
    Variables & Level of Variables &   \\
    \midrule
    Speed limit(km/h) & 30 & 40 & 50\\
    Minimum allowed gap time (s) & 1 & 1.5 & 2\\
    Lane width(m) & 2.5 & 2.75 & 3 \\
    Road Type & One-way & Two-way & Two-way with Median\\
    Breaking levels & 1 & 2 & 3\\
    Arrival Rates(veh/h) & 530 & 750 & 1100\\
    Automation Level & Human Driven & Mixed Traffic & Fully Automated\\
    Time of Day & Day & Night\\ 
    Weather & Clear & Snowy\\
    \bottomrule
  \end{tabular}
  \caption{Variables used to define the experiments}
  \label{table:1} % unique label
\end{table}

\subsection{Simulation and Setup}
\label{sec:2}
The first step of the experiment was to create the simulation with the variables defined in \tref{table:1} using VIRE \cite{farooq2018virtual}, which was implemented in Unity 3D with C$^{\#}$ programming language. The urban environment and 3D models of buildings and street fixtures were designed such that they represent a typical road scene in downtown Toronto. The simulation was projected to an Oculus rift headset with a 15m extended cable. A GSR sensor was synced with the simulation in VIRE to collect VR and stress data simultaneously. An empty experiment room was mapped to the simulation so that the participant can move in it. There was an area cut off to represent the crosswalk with boundaries. A questionnaire was created along with a document with all the experiment possibilities.

\subsection{Data Collection Process}
\label{sec:2}
A variety of participants with different experience levels were needed for this experiment. There were a total of four locations in which the participants were collected, the locations were Ryerson University, City Hall (Toronto, Canada), North York Civic centre, and Markham public library. The participant population at Ryerson University consisted of mostly students. Due to the location of the university most of the participants commuted, and had more experience with cross walks. In City Hall and North York Civil Center, the participants mostly consisted of professionals which ranged from young professionals to seniors. In Markham, families visiting the library participated in the experiments. With the collection of a variety of participants the data were very diverse. 

\subsection{Experiment Details}
\label{sec:2}

\begin{figure}[!h]
  \centering
  \includegraphics[width=0.4\linewidth]{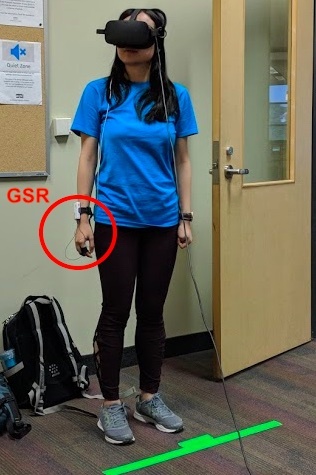} 
  \caption{Participant wearing GSR sensor and VR headset}
  \label{fig:VRGSR} % unique label
\end{figure}

\begin{figure}[!h]
  \centering
  \includegraphics[width=0.475\linewidth]{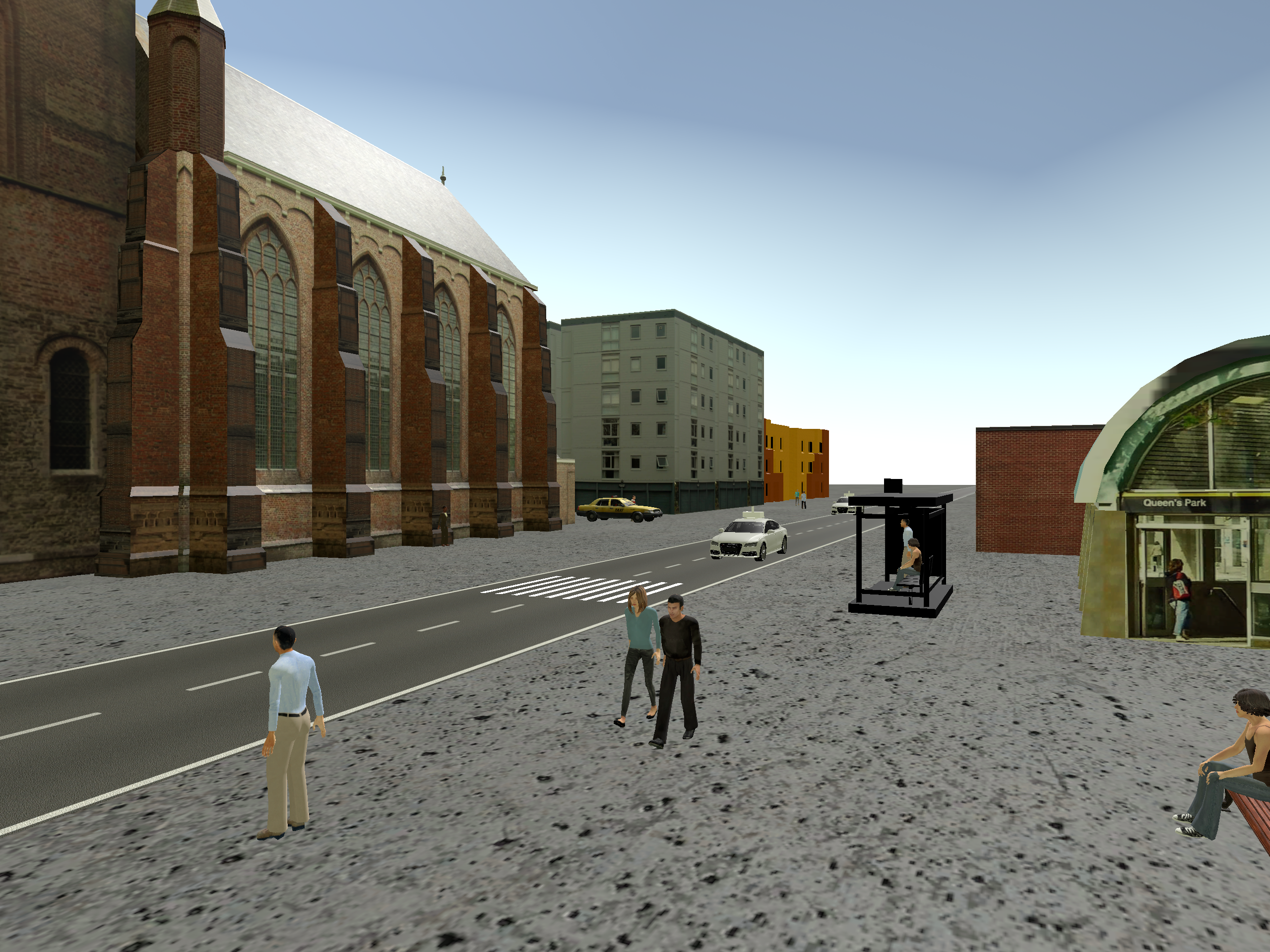}  
    \includegraphics[width=0.475\linewidth]{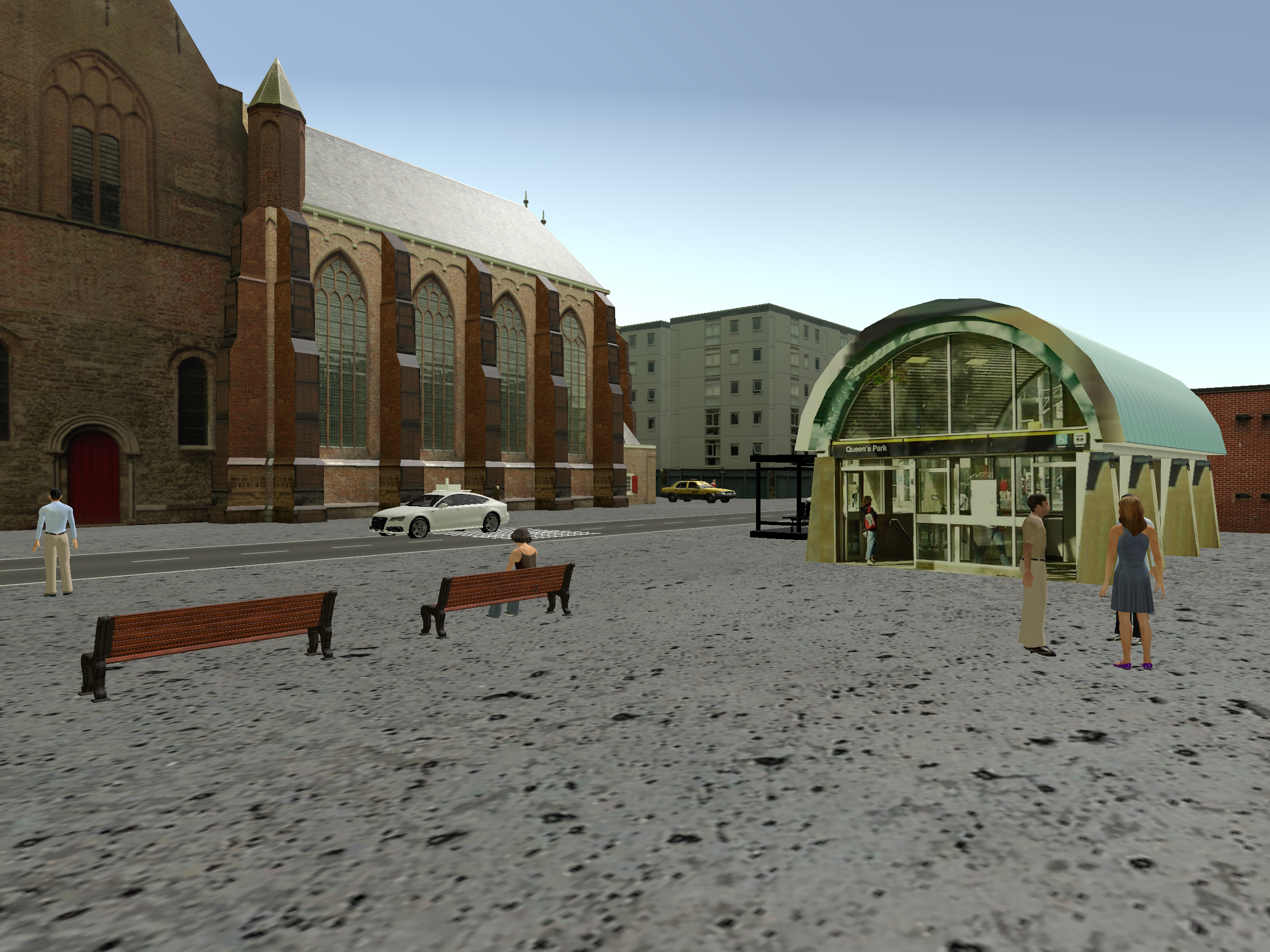}
  \caption{Street scene in VIRE}
  \label{fig:VR2} % unique label
\end{figure}

The experiment started off with the participants filling out the questionnaire about their experience with crossing walks and their sociodemographic information. The participant was then asked to place the GSR sensor on their fingers. The GSR sensor uses sweat to record the stress levels. The sensor sends an electrical signal which became conductive if there was stress on the hand. The more emotionally aroused the participant was the greater the stress levels as there would be a greater amount of sweat on the participant's hand. The participant was asked to equip themselves with the VR helmet and complete a test run of the simulation. To have enough repetitions, each participant was involved in 15 experiments that lasted 20 to 30 seconds each. These experiments were chosen uniformly from the 86 unique combinations outlined by the D-Optimal design. After the experiments were completed, participants were asked to complete a survey to describe their experience and voice their feedback. A participant is shown in \fref{fig:VRGSR} wearing the VR helmet and GSR sensor. \fref{fig:VR2} shows the street scene in VIRE where the experiments were taking place.

% figure example

% figure example

The GSR sensor data were extracted into a comma separated values (csv) file format. This format included a lot of different types of values such as skin conductance and skin resistance being the main focus. Each stress reading was time stamped to the nearest 0.1 second. The skin conductance was indicated to be the stress levels as the conductance of the electrical signals were passed and evaluated using the electrical conductance. One comma separated value file was given for each participant. Due to every participant completing multiple experiments, experiments were divided in the csv file. With this division each experiment was looked into greater depth and stress levels were monitored. The x,y,z coordinated were extracted from the tracking device and were compared to the GSR sensor data. When the stress levels and time stamps were matched, the x,y, and z coordinates were used to calculate speed and acceleration. 

\subsection{Processing of Outcome}
\label{sec:2}
Due to the use of multiple forms of equipment, the collected data were in various formats and units. The VIRE simulator \cite{farooq2018virtual} was recording the x, y, and z axes, every one-tenth of a second. The raw x, y, and z values were first used to calculate the distance $d$ covered by the pedestrian in one-tenth of a second using Pythagoras theorem in \eref{eq:1}. This value was not directly used as the time interval of one-tenth second was too small and may result in unnecessary noise in the calculation of speed and density.

\begin{equation}\label{eq:1}
d = \displaystyle \sqrt{(x_2-x_1)^2+(y_2-y_1)^2+(z_2-z_1)^2}
\end{equation}

Instead, we used one second as out basic unit of time and summed $d$. values for each second \eref{eq:avdis}, where $n=10$.  
\begin{equation}\label{eq:avdis}
{D} = \displaystyle \sum_{i=1}^{n} ({d})
\end{equation}

Once the average distance was obtained for each second ($\Delta T$), it was used to calculate speed and acceleration. The equations for speed and acceleration are shown in \eref{eq:3} and \eref{eq:4}. Where $V$ is velocity, $\Delta T$ is the time interval of one second, and $A$ is acceleration. 

\begin{equation}\label{eq:3}
V =  \frac{D}{\Delta T}
\end{equation}
\begin{equation}\label{eq:4}
A = \frac{\Delta V}{\Delta T}
\end{equation}

In addition to speed and acceleration, we also calculated the cumulative distance travelled $DC$ for each time interval. The cumulative values were used to in the comparative analysis of stress. The equation for cumulative distance is shown in \eref{eq:5}. 

\begin{equation}\label{eq:5}
DC_{t} = \sum_{i=1}^t D_i
\end{equation}

Using the GSR sensors, stress measurements were obtained throughout the length of the experiment. The raw measurements from GSR sensors are intrinsically noisy and require smoothing operation. We tested several smoothing techniques and settled on the Gaussian smoothing as it provided the best fit. The Gaussian smoothing could be used in 2-D and 3-D convolution operators. Most commonly, Gaussian smoothing is used to remove details and noise in an image, but it can also be used for data sets to remove excess noise. The equation for 1-D Gaussian distributions is shown in \eref{eq:6}. 
\begin{equation}
G(x) = e^{-x^2/2{\sigma}^2} \times \frac{1}{\sqrt{2\pi}} 
\label{eq:6}
\end{equation}

\begin{figure}[!h]
  \centering
  \subfigure[]{\includegraphics[width=0.495\linewidth]{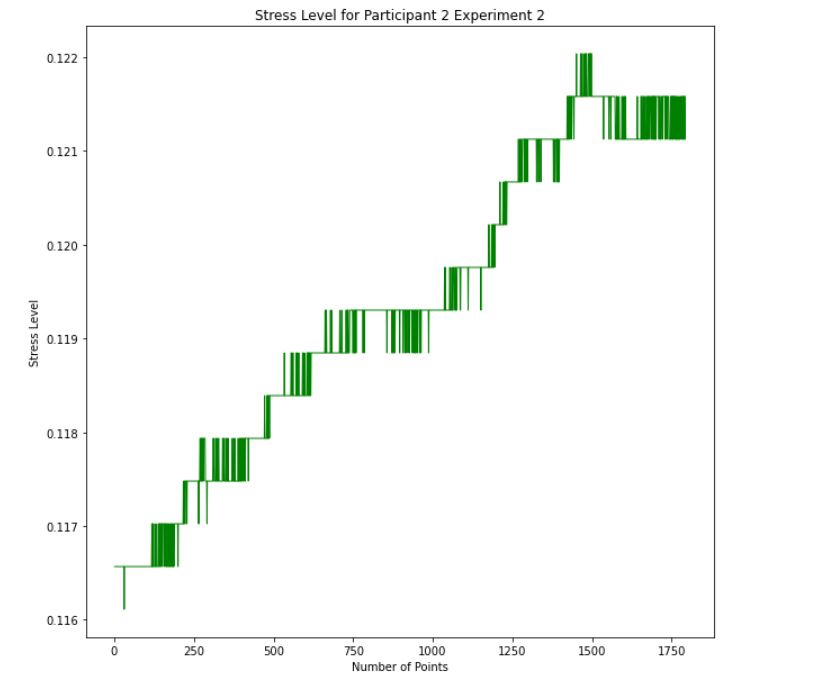}}
  \subfigure[]{\includegraphics[width=0.47\linewidth]{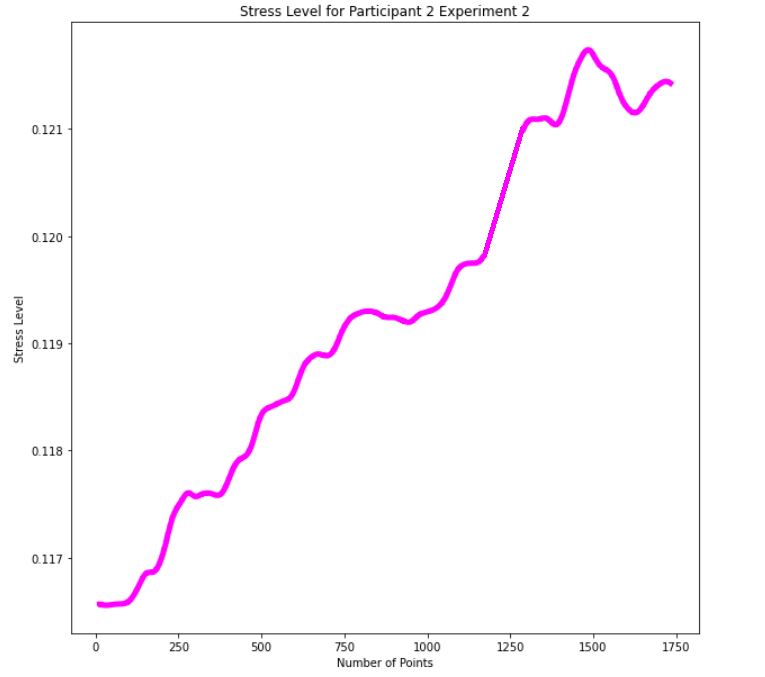}}
  \caption{Gaussian Smoothing and Stress Level (a) and (b)}
  \label{fig:gaussians} % unique label
\end{figure}

\begin{figure}[!h]
  \centering
  \includegraphics[width=0.8\linewidth]{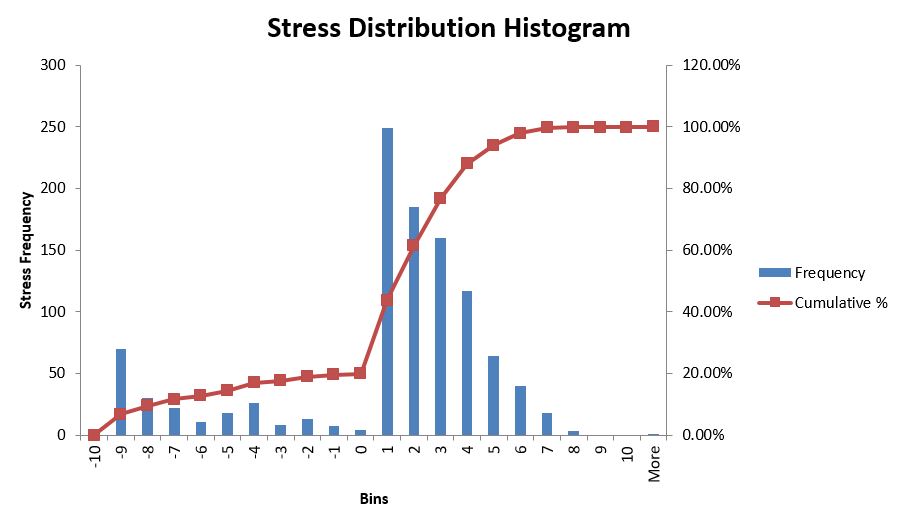} 
  \caption{Stress Distribution Histogram for All Participants and Experiments}
  \label{fig:stressdis} % unique label
\end{figure}

For this equation $\sigma$ is the standard deviation of the distribution and it is assumed with Gaussian distribution that the mean is centered around 0. For our stress levels the Gaussian distribution was centered depending on the location of the peaks in the stress. The data that was collected fitted the 1-D Gaussian distribution. Using python, a code for a Gaussian filter was used to smoothing the stress curves. An example of the Gaussian filtering for one of the participants is shown in \fref{fig:gaussians}. In \fref{fig:gaussians} (a), the original stress levels for Participant 2 Experiment 2 are shown. For \fref{fig:gaussians} (b) the Gaussian effect of smoothing is shown. As seen the smoothing had greatly reduced the noise for the stress levels and is now convenient to analyze. 
The individual stress was also evaluated for each experiment based on the mean stress level. This is due to mean stress levels having a better indication of the overall stress of the experiment, rather than maximum or minimum stress levels. These stress levels were collected and placed into bins to determine the frequency of each mean stress. The mean stresses were rounded based on the closest whole value, due to each individual's stress being a specific value, which cannot be replicated. As shown in \fref{fig:stressdis}, the frequency of stress levels are shown as well as the cumulative percentage of stress. Based on the cumulative percentage, low, medium and high stress levels were labelled for each participant and experiment. Low stress levels were between 0 to 20\% of the cumulative percentage, medium stress levels  were from 20\% to 80\% of the cumulative percentage. The high stress levels included the cumulative percentage of 80\% to 100\%. The stress levels were divided into three levels. Typically cluster analysis could have been used in order to divide into level, but for the scope of this paper the stress levels were divided using a 20-60-20 split. 

% figure example

\section{Results and Discussion}
\label{sec:2}
The section is organized into two parts. First an aggregate level correlation analysis was performed based on the distribution of the variables and the average stress level of the participants per experiment. All participants were included and all variables were taken into consideration. It then present the microscopic and temporal results and analyses of the stress values for several participants and for different experiments. 
\subsection{Correlation Analysis}
\label{sec:2}
A correlation heat map was created with the data collected from the experiments. Here we pooled the data for all participants. Majority of the variables described in the experiments were recorded using a binary system. Since there were multiple combinations of experiments the binary system kept track of which experiment each participant had completed. Each participant has completed on average 10-15 variations of experiments, within the allocated time. One experiment was completed over the course of 15 to 25 seconds. Some participants were able to complete multiple experiments with the same conditions, which made it convenient to compare the results and stress levels. The reaction for each individual to a situation is different, thus when comparing participant stress levels for the same experiment, this should be taken into account. The dependent variable in the experiments was the stress levels, and the independent variables were the sociodemographic and environmental variables.

\begin{figure}
  \centering
  {\includegraphics[width=1\linewidth]{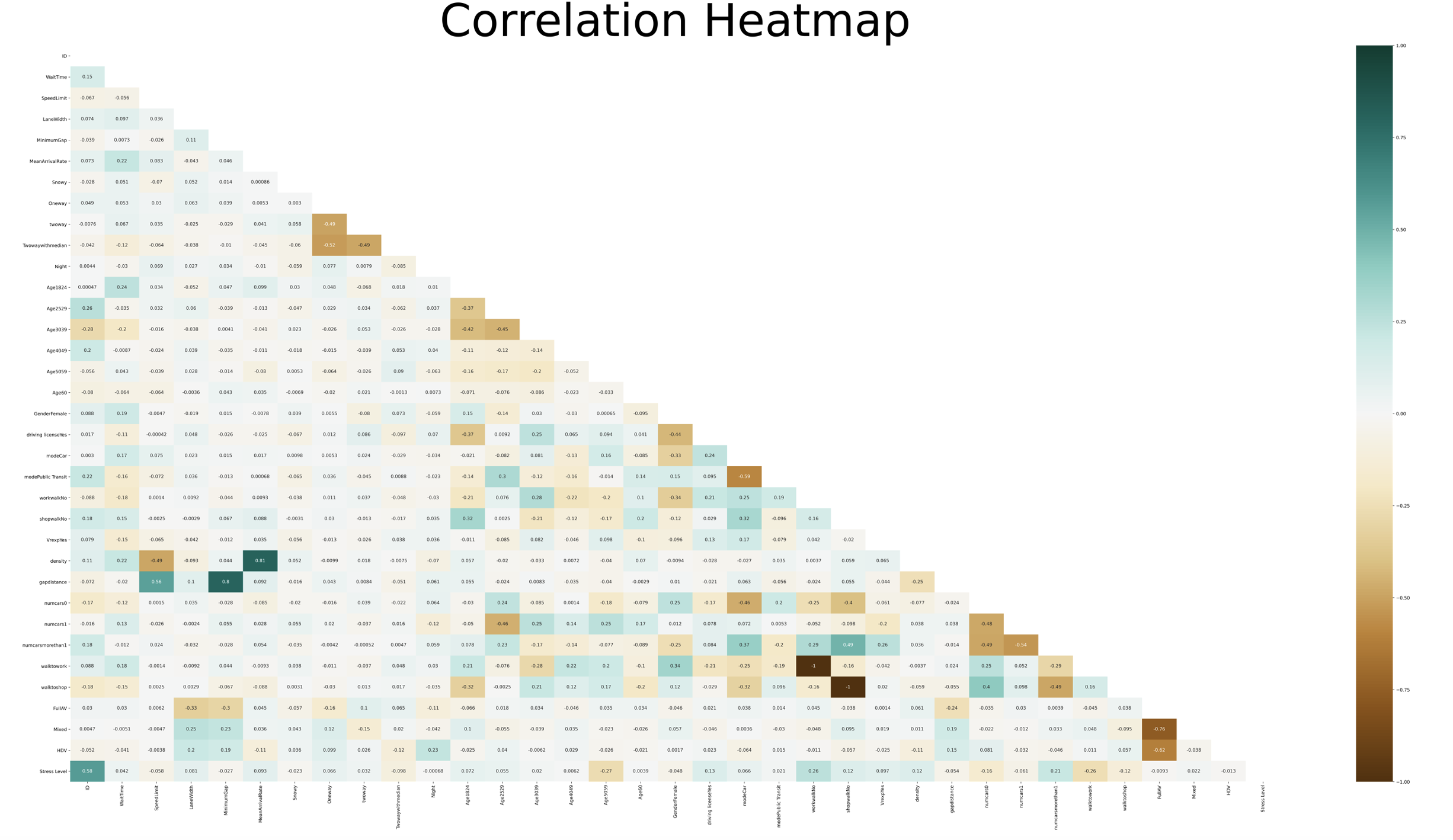}}
  \caption{Correlation HeatMaps for Variables and Stress Levels}
  \label{fig:corrheatmap} % unique label
\end{figure}

\fref{fig:corrheatmap} shows the correlation heat map for all the variables in the experiment. This heat map used the average stress for each experiment conducted for each individual. The stress levels are categorized into 3 levels i.e., low, medium and high levels of stress. The speed limit is in 3 levels as well which are 30, 40 and 50 km/h respectively. The minimum gap time is a value 1, 1.5 and 2 seconds. There are two weather conditions described in this experiment, one of which is snowy and the other which is clear (clear weather conditions). The weather condition described would be a binary value (either 1 or 0). This experiment also contained three different road types which were either two way traffic, one way traffic or two way traffic with median. The day time was also taken into account, either the experiment settings were conducted during the day or the night. The age range of the participants were also monitored to understand the effects of age on the crossing behaviour. If a participant had a driving licence or their main mode of travel was transportation that was considered a variable for the experiment as well. It was asked to the participant if they walked when they went shopping and that answer was displayed as a binary value. The number of cars owned were also asked the groups were if one car was owned, one car or more than one car was owned. The experiment which was conducted determined the level of automation as well, whether the traffic was completely human driven vehicles, mixed traffic or completely autonomous vehicles. 

From the correlation heat map, it is shown that owning more than one car caused a positive correlation with stress levels. This means that if a participant owned more than one car the stress levels for that participant were higher than the average participant. Stress level also has a strong positive correlation with gap distance. The greater the gap distance in the experiment, the higher the stress level for the participants. For negative correlation, the stress levels are higher for participants ages 50 to 59. There is negative correlation between the stress level and if the participant walks to work. This is a logical conclusion as when a participant walks to work, they have more experience with crossing unsignalized intersection, thus they would have less stress. As shown in \fref{fig:corrheatmap}, the correlation between stress level and owning more than one car is 0.21, which in comparison to other variables in stress level is high. Between stress level and walk to work, the correlation is -0.26 which shows a strong negative correlation in comparison to the other correlation values for stress levels.

% figure example

\subsection{Microscopic and Temporal Analysis of Stress}
\label{sec:2}

\begin{figure}
  \centering
  \subfigure[]{\includegraphics[width=0.475\linewidth]{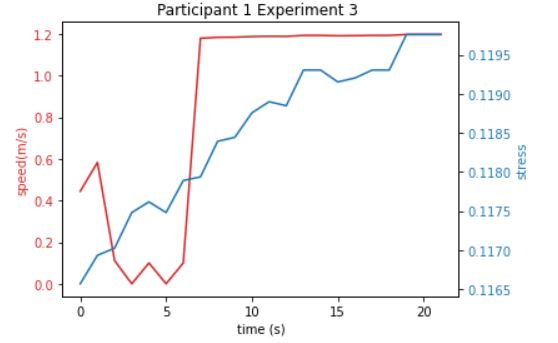}}
  \subfigure[]{\includegraphics[width=0.505\linewidth]{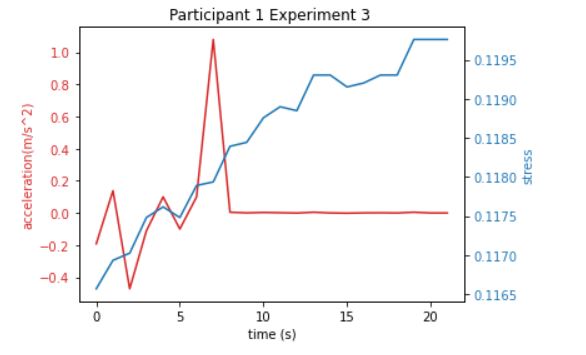}}
  \subfigure[]{\includegraphics[width=0.505\linewidth]{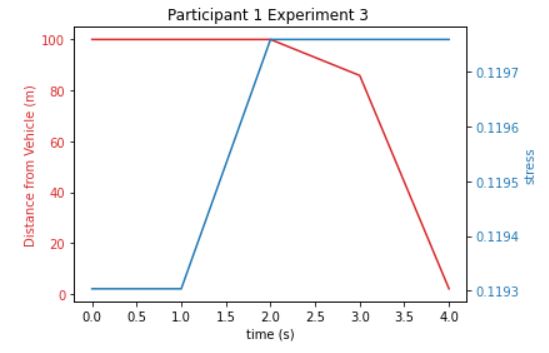}}
  \caption{Participant 1 Experiment 3 Stress, Speed, Acceleration and Distance from Vehicle (a), (b) and (c)(Experiment combination 83, for the variables of speed limit of 40 km/h, lane width of 3m, the road type was a two way road, experiment was conducted at night during snowy weather, and only human driven vehicles)}
  \label{fig:p1e3} % unique label
\end{figure}

\begin{figure}
  \centering
  \subfigure[]{\includegraphics[width=0.495\linewidth]{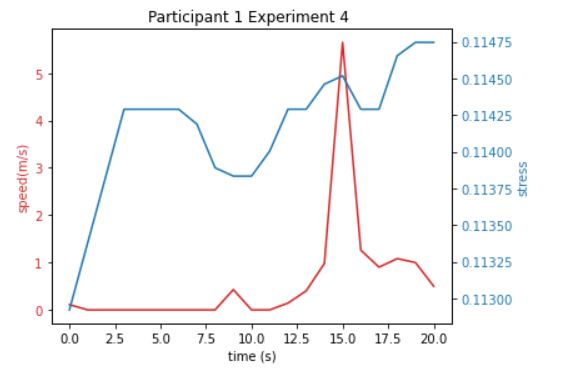}}
  \subfigure[]{\includegraphics[width=0.495\linewidth]{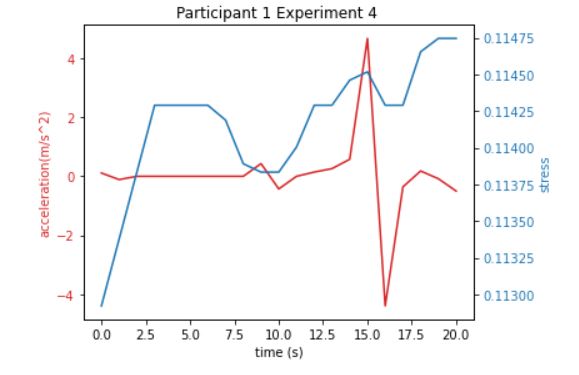}}
  \subfigure[]{\includegraphics[width=0.495\linewidth]{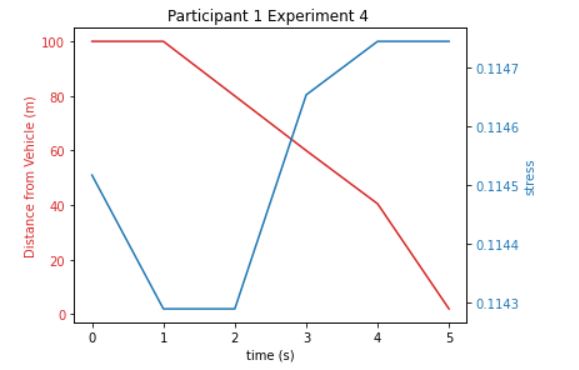}}
  \caption{Participant 1 Experiment 4 Stress, Speed, Acceleration, and Distance from Vehicle (a), (b), and (c)(Experiment combination 83 for the variables of speed limit of 40 km/h, lane width of 3m, the road type was a two way road, experiment was conducted at night during snowy weather, and only human driven vehicles)}
  \label{fig:p1e4} % unique label
\end{figure}

In this section we focused on the data from three participants to get a better understanding of the individual stress levels and connections between crossing behaviour and stress levels. The participants were chosen based on the similar sociodemographic traits and experiments they performed. All the selected participants completed an experiment several times, which enabled us to evaluate the inter and intra participants behaviour and stress measurements. We used speed, acceleration, stress, and distance from an incoming vehicle to analyze the data. \fref{fig:p1e3} shows these indicators for participant 1, while indicators for participant 2 are shown in \fref{fig:p1e4}. The experiment both participants completed was experiment number 83 out of 86 possible experiments. The variable values for this experiment included the speed limit of 40 km/h, lane width of 3m, the road type was a two way road, experiment was conducted at night during snowy weather, and only human driven vehicles were in the flow for these experiments. 

As shown in the \fref{fig:p1e3} for the first participant in experiment 3, the stress level is continuously increasing steadily throughout the experiment. As the experiment was continued the stress level  increased. When the experiment reached 14 seconds the stress drops slightly. Overall for this experiment the participant has an increased amount of stress over time. Speed was calculated throughout the experiment using the equation in \eref{eq:3}(a). The speed increased slightly before drastically decreasing. After that as shown in \fref{fig:p1e3}, the speed was slowly increasing and decreasing before increasing drastically. This suggests to the participant being cautious when crossing the road for this experiment. Once there was a chance to cross, the participant ran across the street. The acceleration versus stress graph shows that the participant crossed the street successfully at the 7$^{th}$ second. After 7 seconds of the experiment the participant has reached the sidewalk on the other side of the crossing. This was known due to the acceleration being 0 and continued to be 0 until the end of the experiment had timed out. It should be noted that for this experiment the respondent's speed is greater than 0 at the beginning of experiment, this was because the previous experiment ended and the participant was still moving a bit. In \fref{fig:p1e3} (c), the distance from the pedestrian to the closest on coming vehicle is shown. As seen, the stress levels increased when the vehicle was closer to the pedestrian, once the vehicle was shown to be further away the stress levels decrease. 

As shown in \fref{fig:p1e4}, the same participant had repeated the experiment with the same conditions between experiment 3 and 4. As shown the stress levels are different, despite completing the same experiment. At around 5 seconds since the start of experiment, the stress drastically decreases for 6 seconds. This is due to the vehicle distance to the pedestrian being further away at that time. After that the stress level rose. Based on the participant's trends it was shown that the participant's max stress is less the second time doing the same experiment compared to the first time.  Like the first experiment, the participant waited for an acceptable gap between vehicles before increasing his speed and crossing the road. The speed increased over 0.5m/s. As shown in the diagram the participant was waiting for the ``right'' opening. The acceleration graph were also created to determine their relationship with stress. The acceleration for the experiment shows that the participant accelerated at one point in time before decreasing drastically and stopping. In Figure \fref{fig:p1e4}, it was shown that when the vehicle distance is further away, the stress levels is lower, but once the vehicle approached the stress levels increase. 

\begin{figure}
  \centering
  \subfigure[]{\includegraphics[width=0.495\linewidth]{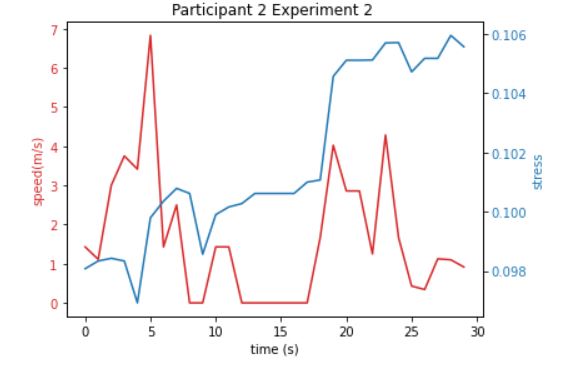}}
  \subfigure[]{\includegraphics[width=0.495\linewidth]{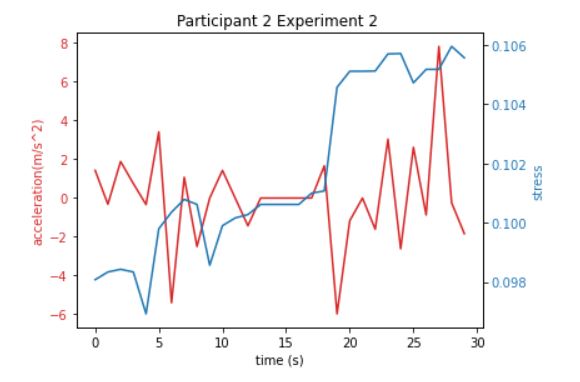}}
  \subfigure[]{\includegraphics[width=0.495\linewidth]{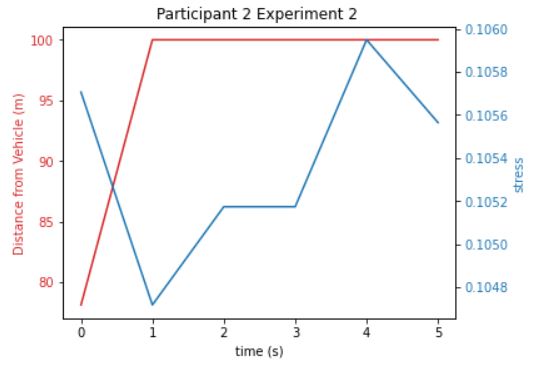}}
  \caption{Participant 2 Experiment 2 Stress, Speed, Acceleration and Distance from Vehicle (a), (b) and (c) (Experiment combination 83, for the variables of speed limit of 40 km/h, lane width of 3m, the road type was a two way road, experiment was conducted at night during snowy weather, and only human driven vehicles)}
  \label{fig:p2e2} % unique label
\end{figure}

For the second participant the stress levels increased drastically, 19 seconds into the experiment. The participant in experiment 2 had the same conditions as the first participant in experiment 3. As shown in \fref{fig:p2e2}, the reaction to stress is different for this participant. This participant did not have a gradual increase in stress, but rather has quick moments in which the stress increased significantly. The speed and method of crossing for this participant is also different from participant 1. The approach for crossing for this person in this experiment is to cross the road in segments. This means the person crossed the first lane when they found an appropriate gap distance between cars. They stopped at that point for a period of time, until they found another acceptable gap, result in they increasing their speed. At the second point the participant increased their speed to 1m/s. Near the end of the experiment the participant reduces their speed and finished their cross. This explains why their stress spikes as, when they planned to increase their speed to cross their stress also increases drastically. The acceleration for this participant varied throughout the experiment. As shown in Figure \fref{fig:p2e2}, after 1 second the distance from the closest vehicle was 100m, which remained until the end of the experiment. The distance from the vehicle to the car is only shown for 5 seconds of the experiment. This is due to the data that was recorded and collected was only analyzed for the last seconds of the experiment.

\begin{figure}
  \centering
  \subfigure[]{\includegraphics[width=0.495\linewidth]{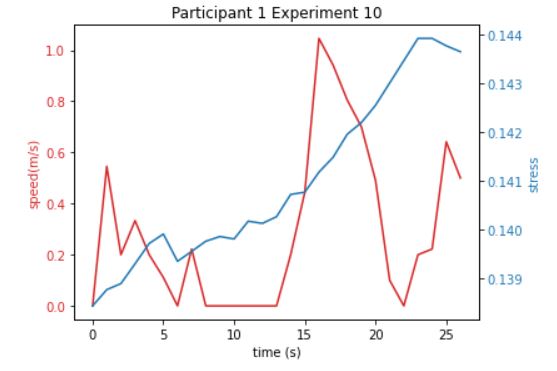}}
  \subfigure[]{\includegraphics[width=0.495\linewidth]{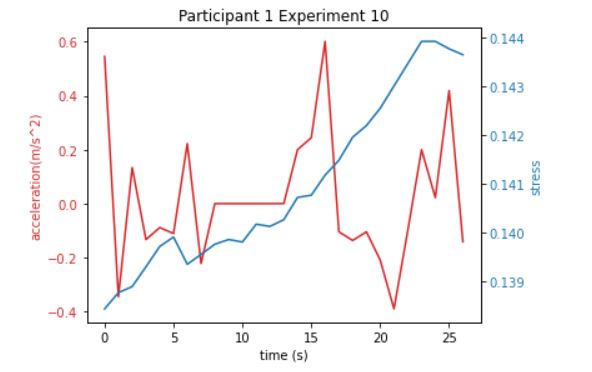}}
  \subfigure[]{\includegraphics[width=0.495\linewidth]{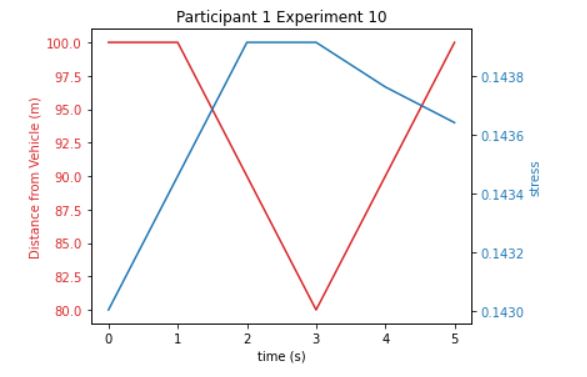}}
  \caption{Participant 1 Experiment 10 Stress, Speed, Acceleration and Distance from Vehicle (a), (b) and (c) (Experiment combination 56,variable conditions of speed limit of 40 km/h, lane width was 2.75, minimum gap was 1 second, it was conducted during the day with clear skies and all vehicles were fully autonomous)}
  \label{fig:p1e10} % unique label
\end{figure}

Participant 1 was evaluated for experiment 10, which had different variable conditions than discussed.  The experiment combination of variables was number 56 out of 86 total experiment combinations.This means that the experiment conditions were as follows,  the speed limit was 40 km/h, lane width was 2.75, minimum gap was 1 second, it was conducted during the day with clear skies and all vehicles were fully autonomous. As shown in \fref{fig:p1e10} (a) and (b) the speed for this experiment raised significantly at 14 seconds. In comparison to other experiments by the same participant, it shows that once an opening was available, the participant used that opening to cross the road. After the speed reached its peak at 1m/s, it dropped to 0m/s before slowly rising and coming to a stop. The stress levels for this experiment continuously roused until the participant reached to the end of the experiment. Compared to the Experiment 3 the participant's stress was greater when interacting within these conditions. This is due to the  result of interaction with fully automated vehicles.

\begin{figure}
  \centering
  \subfigure[]{\includegraphics[width=0.495\linewidth]{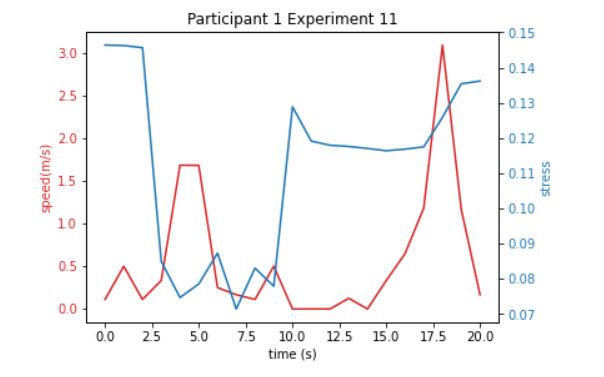}}
  \subfigure[]{\includegraphics[width=0.495\linewidth]{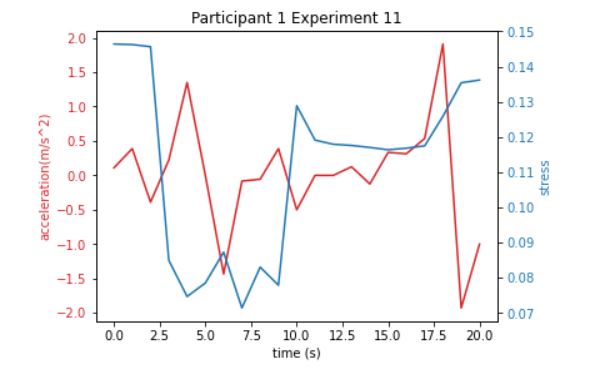}}
  \subfigure[]{\includegraphics[width=0.495\linewidth]{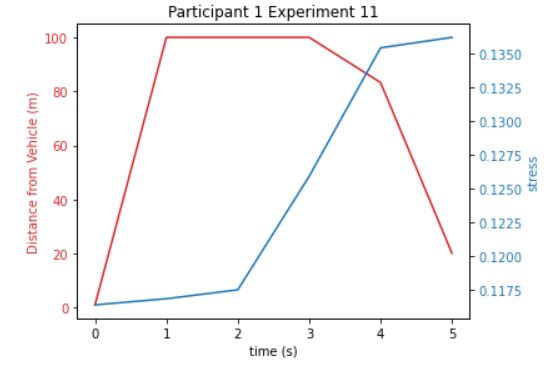}}
  \caption{Participant 1 Experiment 11 Stress, Speed, Acceleration and Distance from Vehicle (a), (b) and (c)(Experiment combination 56,variable conditions of speed limit of 40 km/h, lane width was 2.75, minimum gap was 1 second, it was conducted during the day with clear skies and all vehicles were fully autonomous)}
  \label{fig:p1e11} % unique label
\end{figure}

Participant 1 was evaluated for another experiment with the same variable conditions as experiment 10. This was so that a comparison can be made between the stress levels and speed when the experiment was repeated. As shown in \fref{fig:p1e11} (a) and (b) the speed, stress and acceleration levels were graphed. The stress levels dramatically decreased at 2.5 seconds when the speed was at a local peak. The stress level peaked when the speed was close to zero. Once the participant's speed increased drastically to complete the experiment the stress level also increased. The units of stress are measured in microSiemens (uS). The maximum stress level for this experiment was 0.14uS which was less to the maximum stress level at Experiment 10 (0.144uS). Thus, the exposure to autonomous vehicles a second time, decreased the stress level. Despite this, the stress levels were overall still high and it can be determined that more interaction with autonomous vehicles is needed in order for pedestrians to adapt and communicate with CAVs. 

\begin{figure}
  \centering
  \subfigure[]{\includegraphics[width=0.495\linewidth]{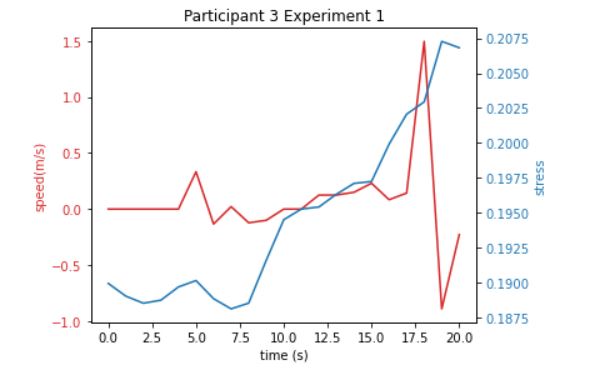}}
  \subfigure[]{\includegraphics[width=0.495\linewidth]{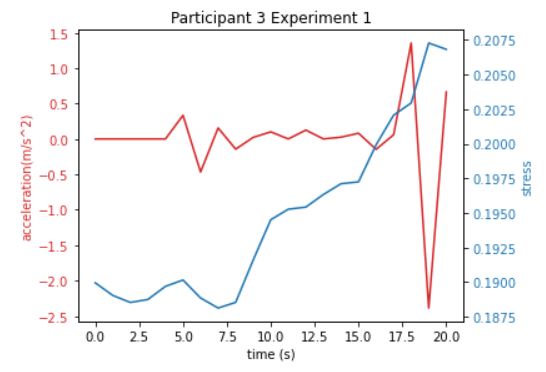}}
  \subfigure[]{\includegraphics[width=0.495\linewidth]{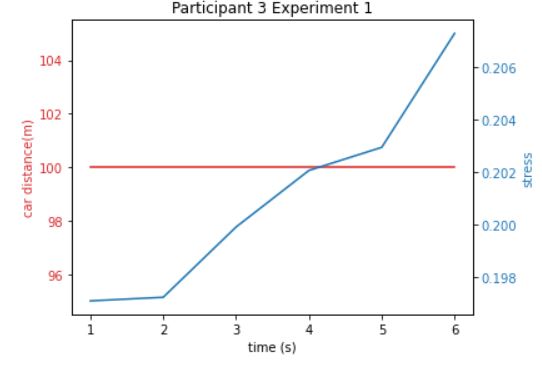}}
  \caption{Participant 3 Experiment 1 Stress, Speed and Acceleration, pedestrian distance to incoming car (a), (b) and (c)(Experiment combination 56,variable conditions of speed limit of 40 km/h, lane width was 2.75, minimum gap was 1 second, it was conducted during the day with clear skies and all vehicles were fully autonomous)}
  \label{fig:p3e1} % unique label
\end{figure}

Participant 3 was evaluated for experiment 1. This experiment had a speed limit of 40km/h, lane with of 2.75m, minimum gap time of 1 second, it was a one way road, taken place during the day when the weather was clear and only fully autonomous vehicles were present. As shown in \fref{fig:p3e1} (a), the speed increased at 6$^{th}$ second for 2 seconds. This is when the pedestrian started to walk but stopped. At 15 seconds the pedestrian walked across the street for 2 seconds before drastically walking back. This could be due to an incoming car. Once the car passed the pedestrian walked quickly across the street and stopped at their destination. The last 6 seconds of the experiment is shown in Figure \fref{fig:p3e1} (c). This indicates that the vehicle is 100 m away, but due to the increasing stress over time for the participant, they still feel stressed until the end of the experiment. This may be due to the adrenaline rush in their body until the task was completed. 

\begin{figure}
  \centering
  \subfigure[]{\includegraphics[width=0.495\linewidth]{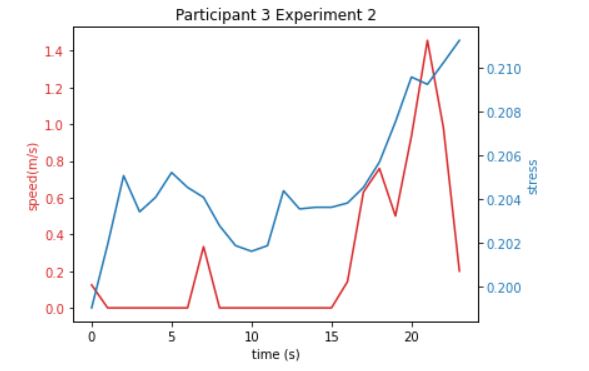}}
  \subfigure[]{\includegraphics[width=0.495\linewidth]{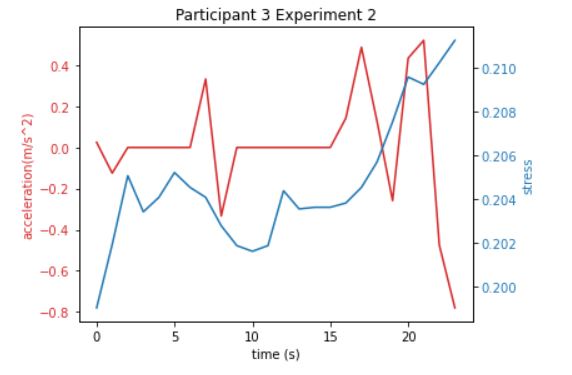}}
  \subfigure[]{\includegraphics[width=0.495\linewidth]{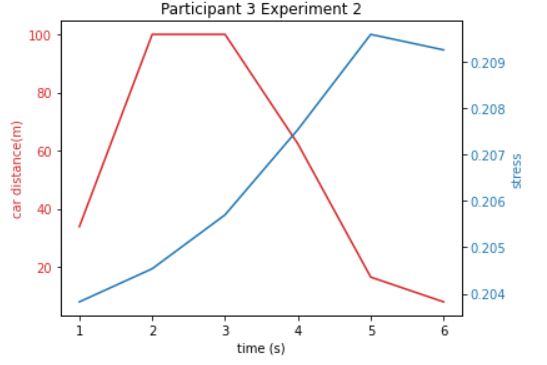}}
  \caption{Participant 3 Experiment 2 Stress, Speed and Acceleration, pedestrian distance to incoming car (a), (b) and (c) (Experiment combination 56, variable conditions of speed limit of 40 km/h, lane width was 2.75, minimum gap was 1 second, it was conducted during the day with clear skies and all vehicles were fully autonomous)}
  \label{fig:p3e2} % unique label
\end{figure}

Participant 3 was also evaluated for experiment 2. For this experiment the speed limit was 40 km/h, lane width was 2.75, minimum gap was 1 second, it was conducted during the day with clear skies and all vehicles were fully autonomous. The experiment combination of the experiment was 56 out of 86. This was the same experiment conditions as Participant 3 experiment 1, participant 1 experiment 10, and participant 1 experiment 11.  As shown in Figure \fref{fig:p3e2} (a), the stress levels decreased at 7 seconds then increased until the end of experiment. The participant's speed increased as the stress levels were increasing. The participants speed decreased at 18 seconds and then increased drastically again until the participant crossed the street successfully. Figure \fref{fig:p3e2}(c) shows the last 6 seconds of the experiment. The stress is increasing as the participant finished crossing the road, as the participant stress levels was increasing once they were in the middle of crossing. 

\begin{figure}
  \centering
  \subfigure[]{\includegraphics[width=0.495\linewidth]{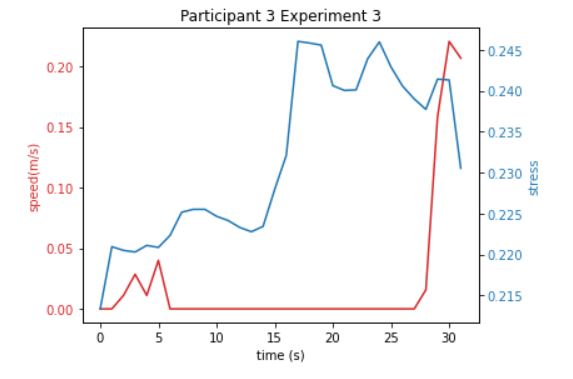}}
  \subfigure[]{\includegraphics[width=0.495\linewidth]{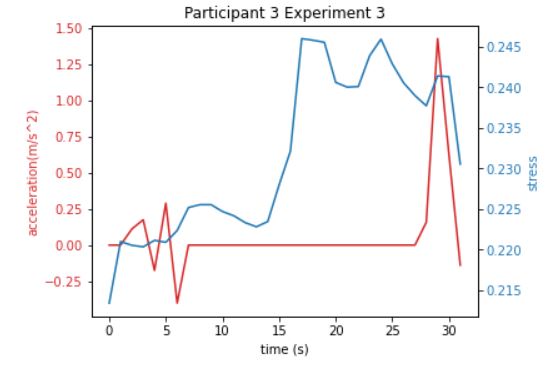}}
  \subfigure[]{\includegraphics[width=0.495\linewidth]{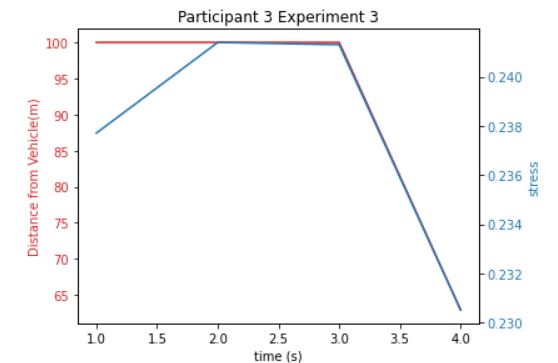}}
  \caption{Participant 3 Experiment 3 Stress, Speed and Acceleration, pedestrian distance to incoming car (a), (b) and (c)(Experiment combination 31, variable conditions of speed limit 50 km/h, lane width  2.75m, minimum gap 1.5 second, the road was one way, it was conducted during the day with clear skies and all vehicles were fully autonomous)}
  \label{fig:p3e3} % unique label
\end{figure}

Participant 3 experiment 3 was evaluated. For this experiment the speed limit was 50 km/h, lane width was 2.75m, minimum gap was 1.5 second, the road was one way, it was conducted during the day with clear skies and all vehicles were fully autonomous. The participant stress levels increased at 14 seconds. The participant's stress was drastically increased until they had crossed the street. The speed of the participant increased and the participant had walked quickly once and opening between vehicles were found. The distance between the car and the pedestrian showed that when the vehicle was far from the pedestrians  the stress level decreased, which is shown in \fref{fig:p3e3} (c).

\section{Conclusion and Future Work}
\label{sec:2}
We used immersive virtual reality and GSR sensor to safely collect and analyze stress of pedestrians when crossing mid-block. The impact of a range of sociedemographic, flow, vehicle type, and geometric variables on the pedestrian stress level is investigated. It was shown that pedestrians have higher stress levels when walking is not their main mode, which may result in the lack of walking experience on urban roads and lack of awareness. Pedestrians were also observed to have slightly higher stress levels initially when interacting with the CAVs. Generally, it was noted throughout all participants that the stress levels were higher when the distance between pedestrians and incoming vehicles decreased. When the vehicles were further, the stress levels were lower. Overall most of the pedestrians had a strategy to walk quickly across the street once the opportunity for them arose. 

We expect that this work will aid in the understanding of pedestrian stress and their reaction to a range of different conditions on the roads in near future. This research is beneficial for stakeholders as when implementing safety features for CAVs, stress levels and reactions of pedestrians can be accounted for. In the transition period from human driven vehicles to CAVs, the stress levels were determined for each scenario and can be used to implement policies and regulations that ensure the safety and comfort of pedestrians on urban roads. Such measures will also result in the acceptance of CAVs on roads by the pedestrians and a sustainable and rapid adoption of this technology.

There are some limitations to this study, which can be addressed in the future work. The number of distractions for pedestrians can be increased to stimulate a more life like environment, such as smartphone use and talking to walking companions. In future work, the analysis of a greater number of participants will be conducted and empirical models will be used to quantify the relationships. The stress levels can be further studied by incorporation different intersection layouts. In future we also plan to study the ability of immersive VR and simulation environment in capturing the true behaviour, stress levels, and decision making abilities of the pedestrians in real life.

\section*{Acknowledgements}
 This research is funded by the Canada Research Chair program, NSERC and Ontario Early Researcher Award. The funding ID is 2017-11-00038.

% BibTeX users use
%\bibliographystyle{cdbibstyle} % mathematics and physical sciences
%\bibliography{ref} % name your BibTeX data base

% Non-BibTeX users please use
%\begin{thebibliography}{1}
%\bibitem{RefJ}
%  % Format for Journal Reference
%  Author: Title. Journal \textbf{Volume}, Pages (Year), \doi{XX.XXXX/XXX}
%\bibitem{RefP}
%  % Format for Proceedings Reference
%  Author: Title. In: Conference, Pages (Year), \doi{XX.XXXX/XXX}
%\bibitem{RefB}
%  % Format for Books
%  Author: Book title. Publisher (Year), \doi{XX.XXXX/XXX}
%\end{thebibliography}

\end{document}